\documentclass[reprint,superscriptaddress]{revtex4-2}
\usepackage[utf8]{inputenc}

\usepackage{amssymb}
\usepackage{amsmath}
\usepackage{graphicx}
\usepackage{colortbl}
\usepackage{xcolor}
\usepackage[caption=false]{subfig}
\usepackage[hidelinks]{hyperref}
\usepackage{comment}
\usepackage[english]{babel}

\usepackage{lineno}

\newcommand*\patchAmsMathEnvironmentForLineno[1]{%
  \expandafter\let\csname old#1\expandafter\endcsname\csname #1\endcsname
  \expandafter\let\csname oldend#1\expandafter\endcsname\csname end#1\endcsname
  \renewenvironment{#1}%
     {\linenomath\csname old#1\endcsname}%
     {\csname oldend#1\endcsname\endlinenomath}}%
\newcommand*\patchBothAmsMathEnvironmentsForLineno[1]{%
  \patchAmsMathEnvironmentForLineno{#1}%
  \patchAmsMathEnvironmentForLineno{#1*}}%
\AtBeginDocument{%
\patchBothAmsMathEnvironmentsForLineno{equation}%
\patchBothAmsMathEnvironmentsForLineno{align}%
\patchBothAmsMathEnvironmentsForLineno{flalign}%
\patchBothAmsMathEnvironmentsForLineno{alignat}%
\patchBothAmsMathEnvironmentsForLineno{gather}%
\patchBothAmsMathEnvironmentsForLineno{multline}%
}

\newcommand{\Dif}{\mathcal{D}}
\newcommand{\mean}[1]{\left\langle #1 \right\rangle}

\newcommand{\Obser}[1]{O \left[ #1\right]}

\newcommand{\Os}[1]{O_s \left[ #1\right]}
\newcommand{\sis}{\text{sign}(s)}
\newcommand{\siS}{\text{sign}(S)}
\newcommand{\siSo}{\text{sign}(S_o)}
\begin{document}
\title{Entropy bound for time reversal markers}
\date{\today}
\author{Gabriel Knotz}
\affiliation{Institute for Theoretical Physics, University of Göttingen, 37077 Göttingen, Germany}
\author{Till M. Muenker}
\affiliation{Third Institute of Physics, University of Göttingen, 37077 Göttingen, Germany}
\author{Timo Betz}
\affiliation{Third Institute of Physics, University of Göttingen, 37077 Göttingen, Germany}
\author{Matthias Krüger}
\affiliation{Institute for Theoretical Physics, University of Göttingen, 37077 Göttingen, Germany}

\begin{abstract}
    We derive a bound for entropy production in terms of the mean of normalizable path-antisymmetric observables. The optimal observable for this bound is shown to be the signum of entropy production, which is often easier determined or estimated than entropy production itself. It can be preserved under coarse graining by use of a simple path grouping algorithm. We demonstrate this relation and its properties using a driven network on a ring, for which the bound saturates for short times for any driving strength. This work can open a way to systematic coarse graining of entropy production.
\end{abstract}
\maketitle


\section{Introduction}

A common way of analyzing complex systems is observation of particle trajectories, e.g., via microscopy 
\cite{squires_fluid_2010, zia_active_2018, bustamante_optical_2021} in biological systems  \cite{fang_nonequilibrium_2020, andrieux_fluctuation_2006,ahmed_active_2015,gladrow_nonequilibrium_2017,hurst_intracellular_2021} or complex fluids \cite{mason_optical_1995,ginot_barrier_2022}.
Detecting and quantifying the deviation from equilibrium, i.e., the violation of detailed balance, based on trajectories is however a challenging task, especially if relevant degrees of freedom are hidden and non-equilibrium processes are random
\cite{roldan_estimating_2010,mehl_role_2012, basu_extrapolation_2018, kahlen_hidden_2018,martinez_inferring_2019, dieball_coarse_2022, van_der_meer_thermodynamic_2022}. Several methods for such detection have been developed. 

The fluctuation dissipation theorem (FDT) connects fluctuations and response functions \cite{kubo_fluctuation-dissipation_1966}, and it is violated out  of equilibrium \cite{agarwal_fluctuation-dissipation_1972, speck_restoring_2006, baiesi_fluctuations_2009}. 
How far away from equilibrium a system is has, e.g., been quantified by using so called effective temperatures or effective energies \cite{martin_comparison_2001, harada_equality_2005,toyabe_nonequilibrium_2010,kruger_fluctuation_2009, ahmed_active_2018,muenker_onsager_2022}.

Another way of detecting broken detailed balance is via entropy production, which has been found to obey a variety of theorems including the  fluctuation theorems  \cite{jarzynski_nonequilibrium_1997,Crooks99,Speck_2007,Seifert2012}. A number of important relations have been found that bound entropy production, such as the thermodynamic uncertainty relation (TUR) \cite{ barato_thermodynamic_2015, Horowitz2020, li_quantifying_2019,van_vu_entropy_2020,dechant_improving_2021, pietzonka_classical_2022,cao_effective_2022, koyuk_thermodynamic_2022}. TUR bounds mean and  variance of currents by entropy production, or vice versa. It has been extended and refined, including path anti-symmetric observables (FTUR) or to more general path weights \cite{dechant_current_2018, hasegawa_fluctuation_2019, dechant_fluctuationresponse_2020, ziyin_universal_2023}. Little is however known about how bounds behave under coarse graining.



In this manuscript we derive an entropy bound in terms of the mean of  path-antisymmetric observables, based on an integrated fluctuation theorem. In contrast to TUR and FTUR, it does not involve the variance of the observable. 
We determine the optimal observable, i.e., the observable that maximizes the bound, to be the signum of entropy production, so that a relation between entropy production and its sign appears. As this relation saturates for a binary process at short times, we argue that no better relation between entropy production and its sign can exist with the same range of validity. The sign of entropy production, and hence the bound for entropy production, can be preserved under coarse graining with a simple path grouping rule. 
We apply these results for a discrete network on a ring. For this network, the signum of entropy production is coarse grained under preservation to the signum of traveled distance, demonstrating how a bound for microscopic entropy production is obtained from a macroscopic observable. Under such coarse graining, entropy production can at most reduce to the original bound. 


\section{Setup and Fluctuation Theorem}
Consider a path observable $\Obser{\omega}$, with path $\omega$ in phase space, with path probability $p[\omega]$, and the average formally given by the sum over paths \cite{altland_condensed_2010}
  $  \mean{O} = \int \Dif \omega ~ \Obser{\omega} p[\omega]$.
To construct a marker for path reversal, the sum is reordered \cite{garcia-garcia_unifying_2010},
\begin{align}
    2\mean{O} = \int \Dif \omega~  \Obser{\omega} p[\omega] + \Obser{\theta \omega} p[\theta \omega].\label{eq:2}
\end{align}
We introduced notation for path reversal, $\theta \omega$, including reversal of time as well as of kinematic reversal of momenta \cite{spinney_entropy_2012}. Validity of Eq.~\eqref{eq:2} requires the sum of paths to include  $\theta \omega$ for any included $\omega$. Adding a zero yields
\begin{align}
    \label{eq:part}
    \begin{split}
    2\mean{O} = \int \Dif \omega\bigl\{ (&\Obser{\omega} + \Obser{\theta \omega}) p[\theta \omega]\\ 
       +  &\Obser{\omega} ( p[\omega] - p[\theta \omega])\bigr\},
    \end{split}
\end{align}
where the term $\Obser{\omega} + \Obser{\theta \omega}$ in the first line of Eq.~\eqref{eq:part} is the path symmetric part of $O$,
\begin{equation}
    \Obser{\omega} + \Obser{\theta \omega} \equiv 2\Os{\omega}.
    \label{eq:prop-symmetric}
\end{equation}
With detailed balance obeyed,  i.e., $p[\omega] = p[\theta \omega]$, antisymmetric observables average to zero, and 
\begin{equation}
    \mean{O} \overset{\text{d.b.}}{=} \langle O_s\rangle.
    \label{eq:Obs-eq}
\end{equation}
Violations of Eq.~\eqref{eq:Obs-eq} thus indicate the breakage of detailed balance  \cite{harris_fluctuation_2007, zia_probability_2007}. While the symmetric part $O_s$ does not appear in the final result, Eq.~\eqref{eq:s-Bound2} below, for the derivation it is useful to start with $O_s$ finite.



To quantify the path reversal properties of cases that break detailed balance, we introduce the stochastic change in entropy defined as the log ratio of path probabilities \cite{Maes2003, seifert_entropy_2005, fischer_free_2020}
\begin{align}
    s_{} &= \log \frac{p[\omega]}{p[\theta \omega]}.
    \label{eq:def-ent}
\end{align}
For simplicity, we will in the following refer to $s$ as the entropy production, despite some caveats regarding this term \footnote{The entropy defined in Eq.~\eqref{eq:def-ent} corresponds to the total change in entropy in overdamped stationary systems. In underdamped or non-stationary systems the boundary terms differ \cite{fischer_free_2020}.}. 
Inserting $s_{}$ into Eq.~\eqref{eq:part} yields
\begin{align}
    2\mean{O} = 2\langle O_s\rangle  +\int \Dif \omega~ \Obser{\omega} \left( 1 - e^{-s_{}} \right)p[\omega].
\end{align}
Reordering the terms yields a fluctuation theorem including $O$
\begin{equation}
    \label{eq:FT}
    \mean{O \left(1 + e^{-s_{}} \right)} = 2\langle O_s\rangle.
\end{equation}
Eq.~\eqref{eq:FT} may be found equivalently from the so called strong detailed fluctuation theorem \cite{garcia-garcia_unifying_2010}, and has been stated in similar form \cite{harris_fluctuation_2007}. 

\section{Entropy bound}

Equation~\eqref{eq:FT} can be used to find bounds for $s$, and we therefore restrict to positive observables, $O[\omega] \geq 0$. This allows Jensen's inequality \cite{jensen_sur_1906, durrett_probabilitytheory_2019} to be applied for the average $\mean{O\dots}/\mean{O}$, to obtain from Eq.~\eqref{eq:FT},
\begin{align}
    \frac{2\mean{O_s} - \mean{O}}{\mean{O}} = \frac{\mean{O e^{-s_{}}}}{\mean{O}} &\geq e^{-\frac{\mean{O s_{}}}{\mean{O}}}.\label{eq:fl}
\end{align}
As expected from Jensen's inequality, Eq.~\eqref{eq:fl}  saturates for small $s$, as seen by expanding it in this limit,
%
\begin{align}
    \frac{\mean{Oe^{-s}}}{\mean{O}} = 1 - \frac{\mean{Os}}{\mean{O}} + \mathcal{O}(s^2) = e^{-\frac{\mean{Os}}{\mean{O}}} + \mathcal{O}(s^2). \label{eq:exp}
\end{align}
Taking the logarithm of Eq.~\eqref{eq:fl} yields a lower bound for the correlation $\mean{Os}$, \footnote{A similar relation was derived in Ref.~\cite{merhav_statistical_2010}, however, with the left hand side always negative.}
\begin{align}
    \log \left( \frac{\mean{O}}{2\mean{O_s} - \mean{O}} \right) &\leq \frac{\mean{O s_{}}}{\mean{O}}.
    \label{eq:G2nd}
\end{align}
Because the conjugate observable $O^*[\omega] = O[\theta \omega] $ is non-negative if $O$ is non-negative, Eq.~\eqref{eq:G2nd} is also valid for $O^*$. The bounds for $\mean{Os}$ and $\mean{O^* s}$ may thus be added to arrive at a bound for $\mean{sO_s}$, the correlation of $s$ and the symmetric part $O_s$
\begin{align}
    2\mean{sO_s} \geq (\mean{O} - \mean{O^*}) \log \left( \frac{\mean{O}}{\mean{O^*}} \right).
    \label{eq:s-Osym-bound}
\end{align}
Notably, when adding Eq.~\eqref{eq:G2nd} for $\mean{Os}$ and $\mean{O^* s}$, the linear term in Eq.~\eqref{eq:exp} drops out, so that Eq.~\eqref{eq:s-Osym-bound} does in general not saturate for small $s$. As will be discussed below, it saturates for a binary process for short times, for any value of $s$.  

A straight forward way to extract a bound for $\mean{s}$ from Eq.~\eqref{eq:s-Osym-bound} is by considering $O_s=1$, i.e.,  path independent. In order for $O$ to be positive, the antisymmetric part, $2O_a [\omega]=O [\omega]-O[\theta\omega]$,  must be normalized to $\left\vert O_a[\omega] \right\vert \leq 1$. This yields
\begin{align}
    \mean{s} &\geq \mean{O_a} \log \left( \frac{ 1 + \mean{O_a}}{1 - \mean{O_a}} \right)\geq0 
    \label{eq:s-Bound2}.
\end{align}
Eq.~\eqref{eq:s-Bound2} is a main result of this manuscript, a bound for entropy production  $\mean{s}$ in terms of the average of  antisymmetric observable $O_a$. This relation is thus fundamentally different from uncertainty relations, which bound entropy production in terms of mean and variance \cite{hasegawa_fluctuation_2019}. 

The condition of $\left\vert O_a[\omega] \right\vert \leq 1$ may seem to be a strong restriction of validity of Eq.~\eqref{eq:s-Bound2}. However, a bound between $\mean{s}$ and $\mean{O_a}$ can only be useful if $O_a$ is normalizable, i.e., if a maximum value of $\underset{\omega}{\max} |O_a| < \infty$ exists. Whenever this maximum exists, $O_a$ can be normalized to fulfill $\left\vert O_a[\omega] \right\vert \leq 1$. Eq.~\eqref{eq:s-Bound2} is thus applicable for any normalizable antisymmetric observable. We also note that the right hand side of Eq.~\eqref{eq:s-Bound2} is non-negative, so that any nonzero $\mean{O_a}$ yields a positive bound for $\mean{s}$.

Eq.~\eqref{eq:s-Bound2} can be read in two ways: 
 (i) A given $\mean{s}$ yields a bound for how far the mean of (any) $O_a$ can deviate from zero. Using, e.g., a time interval from $-t_0$ to $t_0$, $O_a$ can be the time-moment at which a certain event occurs, which is then bound by $\mean{s}$ via Eq.~\eqref{eq:s-Bound2}. This will be investigated in future work. 
 (ii) A given non-vanishing value of $\mean{O_a}$ yields a lower bound for entropy production. We will analyze this below. 

\begin{figure}
    \centering
    \includegraphics[width=\linewidth]{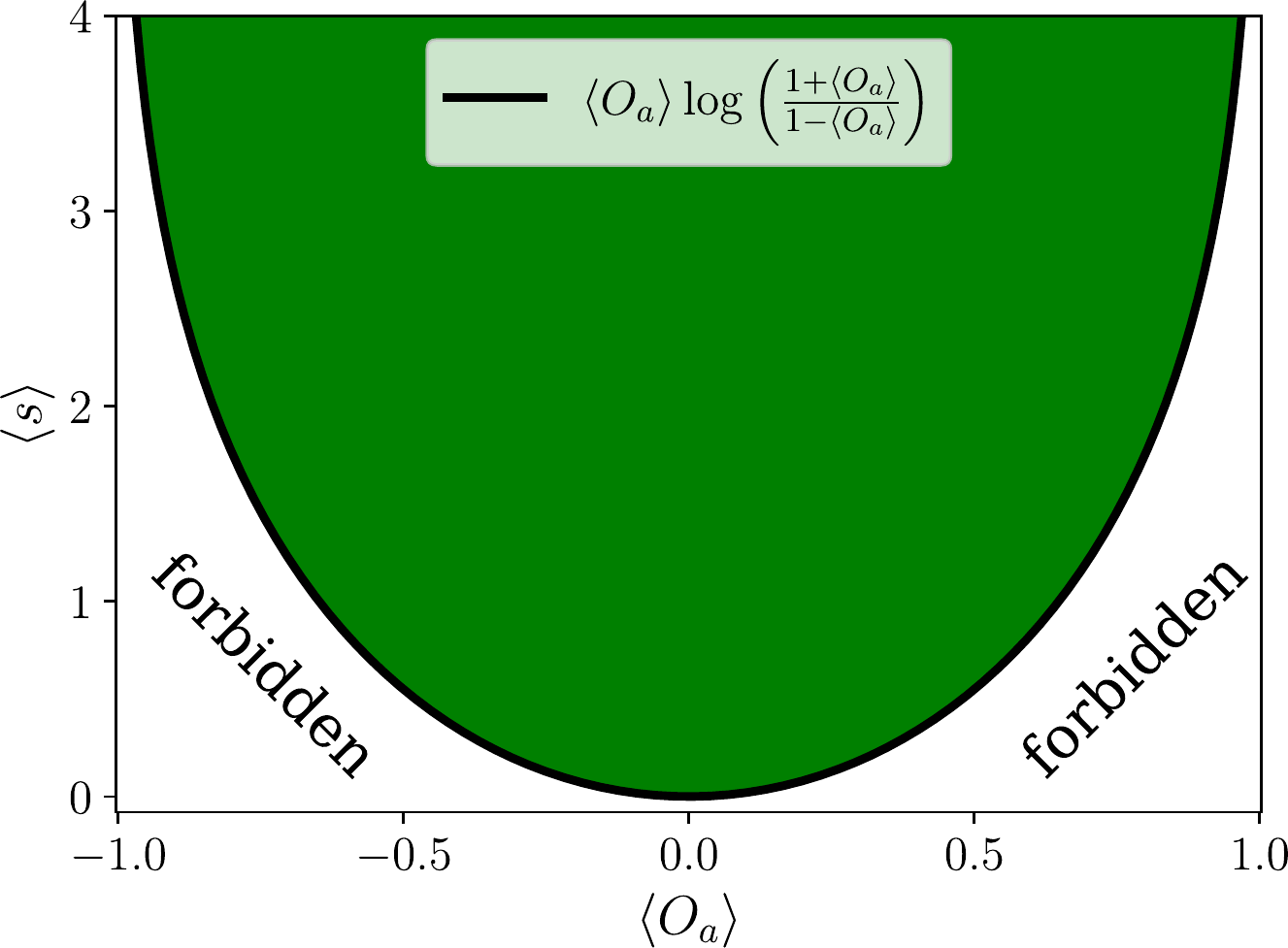}
    \caption{Illustration of Eq.~\eqref{eq:s-Bound2} in terms of  anti-symmetric observable  $\mean{O_a}$, with accessible area marked in green. For $\mean{O_a}\to\pm1$,  the bound diverges logarithmically.}
    \label{fig:bound}
\end{figure}

The form of Eq.~\eqref{eq:s-Bound2} is illustrated in Fig.~\ref{fig:bound}. For small $\mean{O_a}$, the bound grows quadratically in $\mean{O_a}$ while it diverges logarithmically  for $\mean{O_a}\to1$.



\section{Optimal Observable: Signum of Entropy}
Eq.~\eqref{eq:s-Bound2}, as mentioned, is valid for any normalizable antisymmetric observable, and, naturally, the observable that maximizes the right hand side of it, yields the best estimate for $\mean{s}$. Which observable is it? Answering this important question has been found non-trivial for entropy bounds \cite{shiraishi_optimal_2021,dieball_direct_2023}, while it has a clear answer for Eq.~\eqref{eq:s-Bound2}. To see this, rewrite \footnote{Eq.~\eqref{eq:sigs} holds also for $-O_a$ and thus for $|\mean{O_a}|$.} 
\begin{align}
    \mean{O_a} &= \sum_\omega O_a[\omega]p[\omega]
    = \frac{1}{2} \sum_\omega O_a[\omega] (p[\omega] - p[\theta \omega])\notag\\
    &\leq \mean{\text{sign}(s)}.\label{eq:sigs}
\end{align}
In the second step, we used the anti-symmetry of $O_a$. The inequality in the last step of Eq.~\eqref{eq:sigs} follows by noting that the sum is maximized  if $O_a[\omega] = 1$ for $p[\omega] > p[\theta\omega] $ and  $O_a[\omega] = -1$ for $p[\omega] < p[\theta\omega] $. This is the definition of $O_a=\text{sign}(s)$ \footnote{The terms with $p[\omega] = p[\theta\omega]$
 cancel in the sum in Eq.~\eqref{eq:sigs} due to $O_a[\omega]=-O_a[\theta\omega]$, and $O_a[\omega]$ can be chosen arbitrarily in these cases.}.
As the right hand side of Eq.~\eqref{eq:s-Bound2} is a monotonically growing function of $|\mean{O_a}|$ (compare Fig.~\ref{fig:bound}), $O_a=\sis$ yields the optimal bound for $\mean{s}$ from Eq.~\eqref{eq:s-Bound2}. To emphasize this, we write explicitly
\begin{align}
    \mean{s} &\geq \mean{\sis} \log \left( \frac{ 1+\mean{\sis}}{1-\mean{\sis}} \right)\notag\\&\geq \mean{O_a} \log \left( \frac{ 1 + \mean{O_a}}{1 - \mean{O_a}} \right) 
    \label{eq:s-signs}.
\end{align}
The first inequality of Eq.~\eqref{eq:s-signs}  bounds  $\mean{s}$ by $\mean{\sis}$. Writing $\mean{\sis}=\frac{1}{2} \sum_\omega \text{sign}(p[\omega] - p[\theta \omega]) (p[\omega] - p[\theta \omega])$ shows that $\mean{\sis}\geq0$, and that $\mean{\sis}=0$ only if $\mean{s}=0$, i.e.,  Eq.~\eqref{eq:s-signs} yields a finite bound for any finite $\mean{s}$. 
The second inequality of Eq.~\eqref{eq:s-signs} restates that $O_a=\sis$ yields the optimal bound, so that any other $O_a$ lies below it.        
\section{Coarse Graining}
A bound of $\mean{s}$ in terms of $\mean\sis$ is fundamentally interesting, and it is also useful, as, e.g., $\mean\sis$ has beneficial properties under coarse graining. Therefore consider coarse grained paths $\Omega$ with probabilities $P(\Omega)=\sum_{\omega\in\Omega}p(\omega)$, and coarse grained entropy production $S= \log \frac{P[\Omega]}{P[\theta \Omega]}$. Naturally, $O_a=\siS$ fulfills Eq.~\eqref{eq:sigs}, so that, for any grouping of paths
\begin{align}
0\leq\mean{\siS}\leq \mean{\sis}.\label{eq:sb}
\end{align}
Coarse graining thus leads, in general, to a decrease of $\mean{\sis}$, reminiscent of the finding that $\mean{s}$ also decreases under coarse graining \cite{seifert_stochastic_2019}. Notably, grouping paths according to the sign of $s$, i.e., with  ${\rm sign}(s[\omega])={\rm sign}(S[\Omega_o])$ {\it conserves}  $\mean{\sis}$,  
\begin{align}
   \notag \mean{\sis}&=
    \notag \frac{1}{2} \sum_{\Omega_o}\text{sign}(P[\Omega_o] - P[\theta \Omega_o])  \sum_{\omega\in\Omega_o} (p[\omega] - p[\theta \omega])\\
    &= \mean{\siSo}
\end{align}
Under this "optimal" (index o) coarse graining, the bound provided by $\sis$ is invariant, so that the macroscopic $ \mean{\siSo}$ yields the same bound as the microscopic  $\mean{\sis}$. Furthermore, as the bound must hold for $s$ and $S_o$ alike, coarse grained entropy production $S_o$ never falls below the original, microscopic bound. This algorithm thus provides a controlled coarse graining of entropy production, which is especially useful if the bound from $\sis$ is close to $s$. 

\section{Example: Network on a ring}

\begin{figure}
    \centering
    \includegraphics[width=\linewidth]{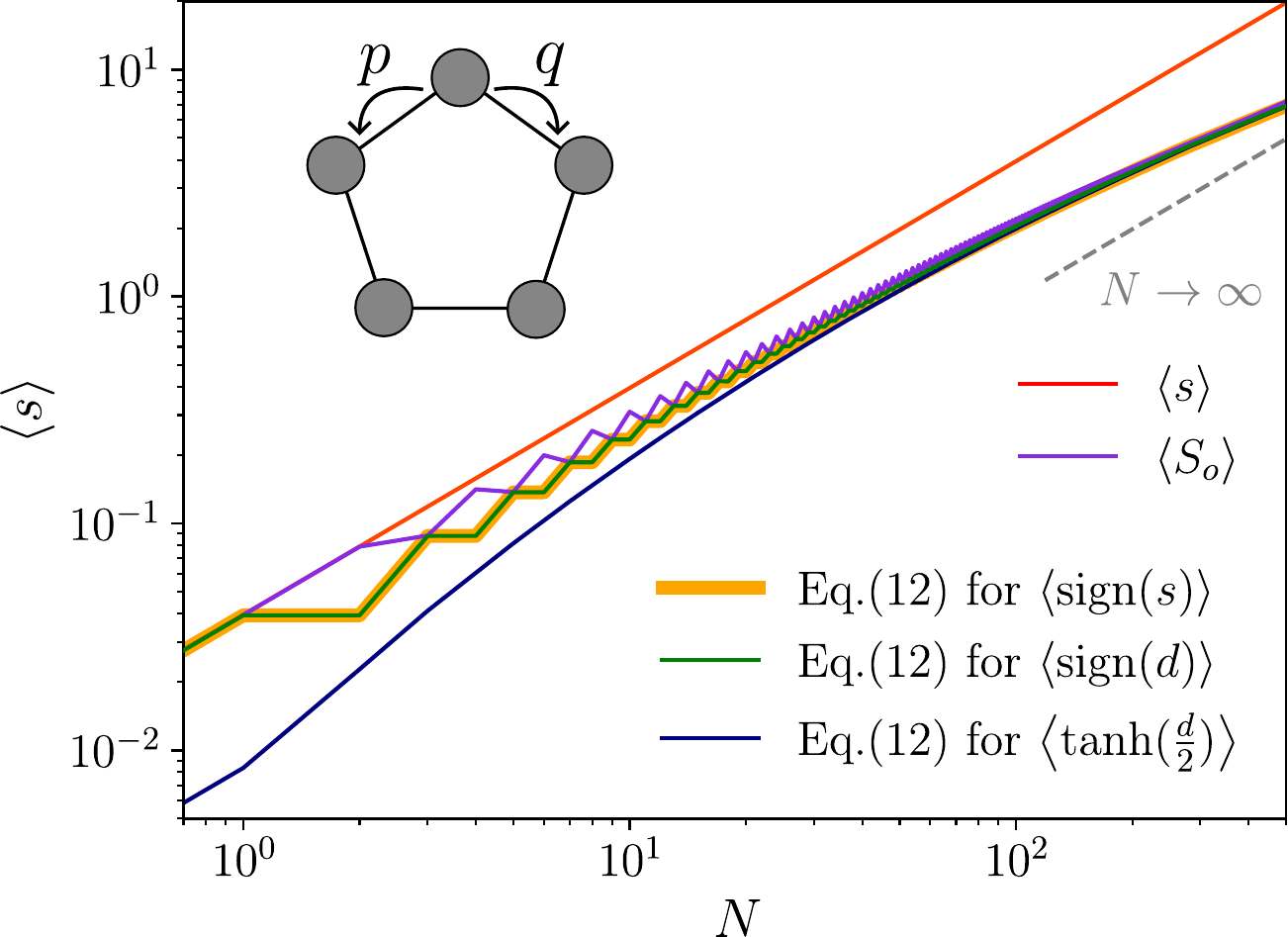}
    \caption{Network on a ring model: Entropy production $\mean{s}$ as a function of $N$, Eq.~\eqref{eq:s}, for $p=0.57$,  versus the bound obtained from Eq.~\eqref{eq:s-Bound2} for various observables as $\sis, \text{sign}(d)$ and $\tanh(\frac{d}{2})$ using numerical simulations. Gray dotted line is the asymtotic limit of large $N$. Graph also shows optimally coarse grained entropy $\mean{S_o}$. }
    \label{fig:two_state}
\end{figure}

To display this in an example,  consider a network on a ring, where every state is connected to two neighbors (see inset sketch of Fig.~\ref{fig:two_state}). In every discrete time step, a particle jumps to the left (right) with probability $p$ ($q$). For $q\not= p$, the system violates detailed balance and shows  a directed flow. After $N$ steps, the probability of finding a specific path with  $n_L$ steps to the left, is given by the binomial distribution
$p[\omega] = \frac{1}{L} p^{n_L} q^{N - n_L}$, with $L$ the number of states in the network. With it, entropy production after $N$ steps is given by,
\begin{align}
    \mean{s} = N(p-q) \log\left( \frac{p}{q} \right).\label{eq:s}
\end{align}
Optimal coarse graining can be performed here in straight forward manner: Because $s=d\log(p/q)$, the sign of entropy production equals the sign of $d = n_L - n_R$ (for  $p>q$), with $n_R$ the number of steps to the right, i.e., $\text{sign}(d[\omega]) = \text{sign}(s[\omega])$. This system thus allows coarse graining towards  measurement of the net displacement $d$, under preservation of the bound. We may expect that $d$ is easier to measure than $s$.

Having established that Eq.~\eqref{eq:s-Bound2} is maximal for $O_a=\text{sign}(d[\omega])$, we can test the quality of the estimate for $\mean{s}$ provided by it.
For $N=1$, $\mean{\text{sign}(d)} = p-q$ and 
\begin{align}
    \mean{s} \overset{N=1}{=} \mean{\text{sign}(d)} \log \left( \frac{1 + \mean{\text{sign}(d)}}{1 - \mean{\text{sign}(d)}} \right).
\end{align}
For $N=1$, the bound of Eq.~\eqref{eq:s-Bound2} thus meets entropy production exactly, for any $p$ and $q$, i.e., arbitrarily far from equilibrium. This is the above mentioned case of binary process, where a particle either jumps right or left. 

Figure \ref{fig:two_state} shows $\mean{s}$ and the bound of Eq.~\eqref{eq:s-Bound2} as a function of $N$. For $N>1$, the bound  grows sublinear in $N$ for an intermediate range, and thus falls below the value of $s$. For $N\gg 1$, it approaches a linear asymptote, which can be found via a large deviation principle. We find for  $p>\frac{1}{2}$ and $N\to\infty$, \cite{arratia_tutorial_1989}
\begin{align}
\mean\sis=\mean{\text{sign}(d)} \sim 1 - \frac{2}{1 - \frac{q}{p}} \frac{\left( \frac{1}{4pq} \right)^{-\frac{N}{2}}}{\sqrt{\frac{\pi}{2} N}} ,\label{eq:signd}
\end{align}
i.e., $\mean{\text{sign}(d)}$ approaches unity exponentially fast with $N$. Because of this, the bound for $s$ of Eq.~\eqref{eq:s-Bound2} grows linear in $N$, and plugging \eqref{eq:signd} into Eq.~\eqref{eq:s-Bound2} yields  $\frac{1}{2} \log \left( \frac{1}{4p q} \right) N$, shown as a gray line in the graph. The ratio between this large $N$ asymptote and $\mean{s}$ of Eq.~\eqref{eq:s} varies between $\frac{1}{2}$ for $p\to1$ and $\frac{1}{4}$ for  $p\to\frac{1}{2}$. 


The coarse graining  groups paths according to their displacement, i.e., $\Omega$ for $d>0$ and $\theta\Omega$ for $d<0$. This way, the coarse grained entropy $\mean{S_o}$ can be determined, which is also shown in Fig.~\ref{fig:two_state}. The curve demonstrates that it, as expected, stays above the bound. As only two coarse grained paths with finite $S_o$ exist
, it is here found from 
\begin{align}
\notag\mean{S_o} &= \log \left( \frac{P[\Omega_o]}{P[\theta \Omega_o]} \right) P[\Omega_o] + \log \left( \frac{P[\theta \Omega_o]}{P[\Omega_o]} \right) P[\theta \Omega_o].\\
&\overset{N {\rm odd}}{=} \mean{\text{sign}(d)} \log \left( \frac{1 + \mean{\text{sign}(d)}}{1 - \mean{\text{sign}(d)}} \right).\label{eq:So}
\end{align}
Notably, the bound Eq. (12) is saturated with respect to $S_o$ for odd $N$ as indicated. For  $N$ even, paths with zero entropy production exist, and the second equality Eq.~\eqref{eq:So} is not valid, 
and Eq. (12) lies below $\mean{S_o}$. For large $N$, these differences vanish, so that the bound of Eq.~\eqref{eq:s-Bound2} and  $\mean{S_o}$ share the same asymptote.   In this system, entropy production may thus be coarse grained by a maximal loss of a factor between $2$ and $4$, depending on $p$, using the optimal algorithm.

According to Eq.~\eqref{eq:sigs}, any other normalizable antisymmetric observable should yield a lower bound, which we examplify by using $O_a=\tanh(d/2)$. Indeed, it lies lower, but approaches the optimal bound for large $N$, because then, a typical trajectory shows $d\gg1$ so that  $\tanh(d/2)$ becomes equivalent to $\text{sign}(d)$.  

While the optimal coarse graining is possible in an exact manner in this model, we expect approximate preservation of $\sis$ to be possible in more complicated systems, which will be investigated in future work. 

\section{Discussion}

Entropy production is bound by the mean of normalizable antisymmetric observables. The optimal observable is identified to be the signum of entropy production, so that we determine a bound between entropy production and its sign, $\sis$. The latter may often be estimated from simple observables, like here, the displacement on a ring. For the investigated network,  $\mean{\sis}$ approaches unity exponentially fast with number of steps, so that the bound grows with the expected linear dependence. Grouping paths according to $\sis$ yields a coarse graining algorithm that preserves $\sis$ and the bound. The presented analysis is not restricted to specific dynamics, and future work may investigate applications to quantum mechanics \cite{esposito_nonequilibrium_2009, landi_irreversible_2021,aron_non_2018}.


\begin{acknowledgments}
We thank \'E. Fodor for a critical reading of the manuscript and Peter Sollich and Cai Dieball for insightful discussions. 
\end{acknowledgments}

\bibliography{references.bib}

\begin{thebibliography}{65}%
\makeatletter
\providecommand \@ifxundefined [1]{%
 \@ifx{#1\undefined}
}%
\providecommand \@ifnum [1]{%
 \ifnum #1\expandafter \@firstoftwo
 \else \expandafter \@secondoftwo
 \fi
}%
\providecommand \@ifx [1]{%
 \ifx #1\expandafter \@firstoftwo
 \else \expandafter \@secondoftwo
 \fi
}%
\providecommand \natexlab [1]{#1}%
\providecommand \enquote  [1]{``#1''}%
\providecommand \bibnamefont  [1]{#1}%
\providecommand \bibfnamefont [1]{#1}%
\providecommand \citenamefont [1]{#1}%
\providecommand \href@noop [0]{\@secondoftwo}%
\providecommand \href [0]{\begingroup \@sanitize@url \@href}%
\providecommand \@href[1]{\@@startlink{#1}\@@href}%
\providecommand \@@href[1]{\endgroup#1\@@endlink}%
\providecommand \@sanitize@url [0]{\catcode `\\12\catcode `\$12\catcode
  `\&12\catcode `\#12\catcode `\^12\catcode `\_12\catcode `\%12\relax}%
\providecommand \@@startlink[1]{}%
\providecommand \@@endlink[0]{}%
\providecommand \url  [0]{\begingroup\@sanitize@url \@url }%
\providecommand \@url [1]{\endgroup\@href {#1}{\urlprefix }}%
\providecommand \urlprefix  [0]{URL }%
\providecommand \Eprint [0]{\href }%
\providecommand \doibase [0]{https://doi.org/}%
\providecommand \selectlanguage [0]{\@gobble}%
\providecommand \bibinfo  [0]{\@secondoftwo}%
\providecommand \bibfield  [0]{\@secondoftwo}%
\providecommand \translation [1]{[#1]}%
\providecommand \BibitemOpen [0]{}%
\providecommand \bibitemStop [0]{}%
\providecommand \bibitemNoStop [0]{.\EOS\space}%
\providecommand \EOS [0]{\spacefactor3000\relax}%
\providecommand \BibitemShut  [1]{\csname bibitem#1\endcsname}%
\let\auto@bib@innerbib\@empty
\bibitem [{\citenamefont {Squires}\ and\ \citenamefont
  {Mason}(2010)}]{squires_fluid_2010}%
  \BibitemOpen
  \bibfield  {author} {\bibinfo {author} {\bibfnamefont {T.~M.}\ \bibnamefont
  {Squires}}\ and\ \bibinfo {author} {\bibfnamefont {T.~G.}\ \bibnamefont
  {Mason}},\ }\href {https://doi.org/10.1146/annurev-fluid-121108-145608}
  {\bibfield  {journal} {\bibinfo  {journal} {Annual Review of Fluid
  Mechanics}\ }\textbf {\bibinfo {volume} {42}},\ \bibinfo {pages} {413–438}
  (\bibinfo {year} {2010})}\BibitemShut {NoStop}%
\bibitem [{\citenamefont {Zia}(2018)}]{zia_active_2018}%
  \BibitemOpen
  \bibfield  {author} {\bibinfo {author} {\bibfnamefont {R.~N.}\ \bibnamefont
  {Zia}},\ }\href {https://doi.org/10.1146/annurev-fluid-122316-044514}
  {\bibfield  {journal} {\bibinfo  {journal} {Annual Review of Fluid
  Mechanics}\ }\textbf {\bibinfo {volume} {50}},\ \bibinfo {pages} {371–405}
  (\bibinfo {year} {2018})}\BibitemShut {NoStop}%
\bibitem [{\citenamefont {Bustamante}\ \emph {et~al.}(2021)\citenamefont
  {Bustamante}, \citenamefont {Chemla}, \citenamefont {Liu},\ and\
  \citenamefont {Wang}}]{bustamante_optical_2021}%
  \BibitemOpen
  \bibfield  {author} {\bibinfo {author} {\bibfnamefont {C.~J.}\ \bibnamefont
  {Bustamante}}, \bibinfo {author} {\bibfnamefont {Y.~R.}\ \bibnamefont
  {Chemla}}, \bibinfo {author} {\bibfnamefont {S.}~\bibnamefont {Liu}},\ and\
  \bibinfo {author} {\bibfnamefont {M.~D.}\ \bibnamefont {Wang}},\ }\href
  {https://doi.org/10.1038/s43586-021-00021-6} {\bibfield  {journal} {\bibinfo
  {journal} {Nature Reviews Methods Primers}\ }\textbf {\bibinfo {volume}
  {1}},\ \bibinfo {pages} {25} (\bibinfo {year} {2021})}\BibitemShut {NoStop}%
\bibitem [{\citenamefont {Fang}\ and\ \citenamefont
  {Wang}(2020)}]{fang_nonequilibrium_2020}%
  \BibitemOpen
  \bibfield  {author} {\bibinfo {author} {\bibfnamefont {X.}~\bibnamefont
  {Fang}}\ and\ \bibinfo {author} {\bibfnamefont {J.}~\bibnamefont {Wang}},\
  }\href {https://doi.org/10.1146/annurev-biophys-121219-081656} {\bibfield
  {journal} {\bibinfo  {journal} {Annual Review of Biophysics}\ }\textbf
  {\bibinfo {volume} {49}},\ \bibinfo {pages} {227–246} (\bibinfo {year}
  {2020})}\BibitemShut {NoStop}%
\bibitem [{\citenamefont {Andrieux}\ and\ \citenamefont
  {Gaspard}(2006)}]{andrieux_fluctuation_2006}%
  \BibitemOpen
  \bibfield  {author} {\bibinfo {author} {\bibfnamefont {D.}~\bibnamefont
  {Andrieux}}\ and\ \bibinfo {author} {\bibfnamefont {P.}~\bibnamefont
  {Gaspard}},\ }\href {https://doi.org/10.1103/PhysRevE.74.011906} {\bibfield
  {journal} {\bibinfo  {journal} {Physical Review E}\ }\textbf {\bibinfo
  {volume} {74}},\ \bibinfo {pages} {011906} (\bibinfo {year}
  {2006})}\BibitemShut {NoStop}%
\bibitem [{\citenamefont {Ahmed}\ \emph {et~al.}(2015)\citenamefont {Ahmed},
  \citenamefont {Fodor},\ and\ \citenamefont {Betz}}]{ahmed_active_2015}%
  \BibitemOpen
  \bibfield  {author} {\bibinfo {author} {\bibfnamefont {W.~W.}\ \bibnamefont
  {Ahmed}}, \bibinfo {author} {\bibfnamefont {{\'{E}}.}~\bibnamefont {Fodor}},\
  and\ \bibinfo {author} {\bibfnamefont {T.}~\bibnamefont {Betz}},\ }\href
  {https://doi.org/10.1016/j.bbamcr.2015.05.022} {\bibfield  {journal}
  {\bibinfo  {journal} {Biochimica et Biophysica Acta (BBA) - Molecular Cell
  Research}\ }\textbf {\bibinfo {volume} {1853}},\ \bibinfo {pages}
  {3083–3094} (\bibinfo {year} {2015})},\ \bibinfo {note}
  {mechanobiology}\BibitemShut {NoStop}%
\bibitem [{\citenamefont {Gladrow}\ \emph {et~al.}(2017)\citenamefont
  {Gladrow}, \citenamefont {Broedersz},\ and\ \citenamefont
  {Schmidt}}]{gladrow_nonequilibrium_2017}%
  \BibitemOpen
  \bibfield  {author} {\bibinfo {author} {\bibfnamefont {J.}~\bibnamefont
  {Gladrow}}, \bibinfo {author} {\bibfnamefont {C.~P.}\ \bibnamefont
  {Broedersz}},\ and\ \bibinfo {author} {\bibfnamefont {C.~F.}\ \bibnamefont
  {Schmidt}},\ }\href {https://doi.org/10.1103/PhysRevE.96.022408} {\bibfield
  {journal} {\bibinfo  {journal} {Physical Review E}\ }\textbf {\bibinfo
  {volume} {96}},\ \bibinfo {pages} {022408} (\bibinfo {year}
  {2017})}\BibitemShut {NoStop}%
\bibitem [{\citenamefont {Hurst}\ \emph {et~al.}(2021)\citenamefont {Hurst},
  \citenamefont {Vos}, \citenamefont {Brandt},\ and\ \citenamefont
  {Betz}}]{hurst_intracellular_2021}%
  \BibitemOpen
  \bibfield  {author} {\bibinfo {author} {\bibfnamefont {S.}~\bibnamefont
  {Hurst}}, \bibinfo {author} {\bibfnamefont {B.~E.}\ \bibnamefont {Vos}},
  \bibinfo {author} {\bibfnamefont {M.}~\bibnamefont {Brandt}},\ and\ \bibinfo
  {author} {\bibfnamefont {T.}~\bibnamefont {Betz}},\ }\href
  {https://doi.org/10.1038/s41567-021-01368-z} {\bibfield  {journal} {\bibinfo
  {journal} {Nature Physics}\ }\textbf {\bibinfo {volume} {17}},\ \bibinfo
  {pages} {1270} (\bibinfo {year} {2021})}\BibitemShut {NoStop}%
\bibitem [{\citenamefont {Mason}\ and\ \citenamefont
  {Weitz}(1995)}]{mason_optical_1995}%
  \BibitemOpen
  \bibfield  {author} {\bibinfo {author} {\bibfnamefont {T.~G.}\ \bibnamefont
  {Mason}}\ and\ \bibinfo {author} {\bibfnamefont {D.~A.}\ \bibnamefont
  {Weitz}},\ }\href {https://doi.org/10.1103/PhysRevLett.74.1250} {\bibfield
  {journal} {\bibinfo  {journal} {Physical Review Letters}\ }\textbf {\bibinfo
  {volume} {74}},\ \bibinfo {pages} {1250–1253} (\bibinfo {year}
  {1995})}\BibitemShut {NoStop}%
\bibitem [{\citenamefont {Ginot}\ \emph {et~al.}(2022)\citenamefont {Ginot},
  \citenamefont {Caspers}, \citenamefont {Krüger},\ and\ \citenamefont
  {Bechinger}}]{ginot_barrier_2022}%
  \BibitemOpen
  \bibfield  {author} {\bibinfo {author} {\bibfnamefont {F.}~\bibnamefont
  {Ginot}}, \bibinfo {author} {\bibfnamefont {J.}~\bibnamefont {Caspers}},
  \bibinfo {author} {\bibfnamefont {M.}~\bibnamefont {Krüger}},\ and\ \bibinfo
  {author} {\bibfnamefont {C.}~\bibnamefont {Bechinger}},\ }\href
  {https://doi.org/10.1103/PhysRevLett.128.028001} {\bibfield  {journal}
  {\bibinfo  {journal} {Physical Review Letters}\ }\textbf {\bibinfo {volume}
  {128}},\ \bibinfo {pages} {028001} (\bibinfo {year} {2022})}\BibitemShut
  {NoStop}%
\bibitem [{\citenamefont {Rold{\'{a}}n}\ and\ \citenamefont
  {Parrondo}(2010)}]{roldan_estimating_2010}%
  \BibitemOpen
  \bibfield  {author} {\bibinfo {author} {\bibfnamefont {{\'{E}}.}~\bibnamefont
  {Rold{\'{a}}n}}\ and\ \bibinfo {author} {\bibfnamefont {J.~M.~R.}\
  \bibnamefont {Parrondo}},\ }\href
  {https://doi.org/10.1103/PhysRevLett.105.150607} {\bibfield  {journal}
  {\bibinfo  {journal} {Physical Review Letters}\ }\textbf {\bibinfo {volume}
  {105}},\ \bibinfo {pages} {150607} (\bibinfo {year} {2010})}\BibitemShut
  {NoStop}%
\bibitem [{\citenamefont {Mehl}\ \emph {et~al.}(2012)\citenamefont {Mehl},
  \citenamefont {Lander}, \citenamefont {Bechinger}, \citenamefont {Blickle},\
  and\ \citenamefont {Seifert}}]{mehl_role_2012}%
  \BibitemOpen
  \bibfield  {author} {\bibinfo {author} {\bibfnamefont {J.}~\bibnamefont
  {Mehl}}, \bibinfo {author} {\bibfnamefont {B.}~\bibnamefont {Lander}},
  \bibinfo {author} {\bibfnamefont {C.}~\bibnamefont {Bechinger}}, \bibinfo
  {author} {\bibfnamefont {V.}~\bibnamefont {Blickle}},\ and\ \bibinfo {author}
  {\bibfnamefont {U.}~\bibnamefont {Seifert}},\ }\href
  {https://doi.org/10.1103/PhysRevLett.108.220601} {\bibfield  {journal}
  {\bibinfo  {journal} {Physical Review Letters}\ }\textbf {\bibinfo {volume}
  {108}},\ \bibinfo {pages} {220601} (\bibinfo {year} {2012})}\BibitemShut
  {NoStop}%
\bibitem [{\citenamefont {Basu}\ \emph {et~al.}(2018)\citenamefont {Basu},
  \citenamefont {Helden},\ and\ \citenamefont
  {Krüger}}]{basu_extrapolation_2018}%
  \BibitemOpen
  \bibfield  {author} {\bibinfo {author} {\bibfnamefont {U.}~\bibnamefont
  {Basu}}, \bibinfo {author} {\bibfnamefont {L.}~\bibnamefont {Helden}},\ and\
  \bibinfo {author} {\bibfnamefont {M.}~\bibnamefont {Krüger}},\ }\href
  {https://doi.org/10.1103/PhysRevLett.120.180604} {\bibfield  {journal}
  {\bibinfo  {journal} {Physical Review Letters}\ }\textbf {\bibinfo {volume}
  {120}},\ \bibinfo {pages} {180604} (\bibinfo {year} {2018})}\BibitemShut
  {NoStop}%
\bibitem [{\citenamefont {Kahlen}\ and\ \citenamefont
  {Ehrich}(2018)}]{kahlen_hidden_2018}%
  \BibitemOpen
  \bibfield  {author} {\bibinfo {author} {\bibfnamefont {M.}~\bibnamefont
  {Kahlen}}\ and\ \bibinfo {author} {\bibfnamefont {J.}~\bibnamefont
  {Ehrich}},\ }\href {https://doi.org/10.1088/1742-5468/aac2fd} {\bibfield
  {journal} {\bibinfo  {journal} {Journal of Statistical Mechanics: Theory and
  Experiment}\ }\textbf {\bibinfo {volume} {2018}},\ \bibinfo {pages} {063204}
  (\bibinfo {year} {2018})}\BibitemShut {NoStop}%
\bibitem [{\citenamefont {Martínez}\ \emph {et~al.}(2019)\citenamefont
  {Martínez}, \citenamefont {Bisker}, \citenamefont {Horowitz},\ and\
  \citenamefont {Parrondo}}]{martinez_inferring_2019}%
  \BibitemOpen
  \bibfield  {author} {\bibinfo {author} {\bibfnamefont {I.~A.}\ \bibnamefont
  {Martínez}}, \bibinfo {author} {\bibfnamefont {G.}~\bibnamefont {Bisker}},
  \bibinfo {author} {\bibfnamefont {J.~M.}\ \bibnamefont {Horowitz}},\ and\
  \bibinfo {author} {\bibfnamefont {J.~M.~R.}\ \bibnamefont {Parrondo}},\
  }\href {https://doi.org/10.1038/s41467-019-11051-w} {\bibfield  {journal}
  {\bibinfo  {journal} {Nature Communications}\ }\textbf {\bibinfo {volume}
  {10}},\ \bibinfo {pages} {3542} (\bibinfo {year} {2019})}\BibitemShut
  {NoStop}%
\bibitem [{\citenamefont {Dieball}\ and\ \citenamefont
  {Godec}(2022)}]{dieball_coarse_2022}%
  \BibitemOpen
  \bibfield  {author} {\bibinfo {author} {\bibfnamefont {C.}~\bibnamefont
  {Dieball}}\ and\ \bibinfo {author} {\bibfnamefont {A.}~\bibnamefont
  {Godec}},\ }\href {https://doi.org/10.1103/PhysRevResearch.4.033243}
  {\bibfield  {journal} {\bibinfo  {journal} {Physical Review Research}\
  }\textbf {\bibinfo {volume} {4}},\ \bibinfo {pages} {033243} (\bibinfo {year}
  {2022})}\BibitemShut {NoStop}%
\bibitem [{\citenamefont {van~der Meer}\ \emph {et~al.}(2022)\citenamefont
  {van~der Meer}, \citenamefont {Ertel},\ and\ \citenamefont
  {Seifert}}]{van_der_meer_thermodynamic_2022}%
  \BibitemOpen
  \bibfield  {author} {\bibinfo {author} {\bibfnamefont {J.}~\bibnamefont
  {van~der Meer}}, \bibinfo {author} {\bibfnamefont {B.}~\bibnamefont
  {Ertel}},\ and\ \bibinfo {author} {\bibfnamefont {U.}~\bibnamefont
  {Seifert}},\ }\href {https://doi.org/10.1103/PhysRevX.12.031025} {\bibfield
  {journal} {\bibinfo  {journal} {Physical Review X}\ }\textbf {\bibinfo
  {volume} {12}},\ \bibinfo {pages} {031025} (\bibinfo {year}
  {2022})}\BibitemShut {NoStop}%
\bibitem [{\citenamefont {Kubo}(1966)}]{kubo_fluctuation-dissipation_1966}%
  \BibitemOpen
  \bibfield  {author} {\bibinfo {author} {\bibfnamefont {R.}~\bibnamefont
  {Kubo}},\ }\href {https://doi.org/10.1088/0034-4885/29/1/306} {\bibfield
  {journal} {\bibinfo  {journal} {Reports on Progress in Physics}\ }\textbf
  {\bibinfo {volume} {29}},\ \bibinfo {pages} {255} (\bibinfo {year}
  {1966})}\BibitemShut {NoStop}%
\bibitem [{\citenamefont
  {Agarwal}(1972)}]{agarwal_fluctuation-dissipation_1972}%
  \BibitemOpen
  \bibfield  {author} {\bibinfo {author} {\bibfnamefont {G.~S.}\ \bibnamefont
  {Agarwal}},\ }\href {https://doi.org/10.1007/BF01391621} {\bibfield
  {journal} {\bibinfo  {journal} {Zeitschrift für Physik A Hadrons and
  nuclei}\ }\textbf {\bibinfo {volume} {252}},\ \bibinfo {pages} {25–38}
  (\bibinfo {year} {1972})}\BibitemShut {NoStop}%
\bibitem [{\citenamefont {Speck}\ and\ \citenamefont
  {Seifert}(2006)}]{speck_restoring_2006}%
  \BibitemOpen
  \bibfield  {author} {\bibinfo {author} {\bibfnamefont {T.}~\bibnamefont
  {Speck}}\ and\ \bibinfo {author} {\bibfnamefont {U.}~\bibnamefont
  {Seifert}},\ }\href {https://doi.org/10.1209/epl/i2005-10549-4} {\bibfield
  {journal} {\bibinfo  {journal} {Europhysics Letters (EPL)}\ }\textbf
  {\bibinfo {volume} {74}},\ \bibinfo {pages} {391–396} (\bibinfo {year}
  {2006})}\BibitemShut {NoStop}%
\bibitem [{\citenamefont {Baiesi}\ \emph {et~al.}(2009)\citenamefont {Baiesi},
  \citenamefont {Maes},\ and\ \citenamefont
  {Wynants}}]{baiesi_fluctuations_2009}%
  \BibitemOpen
  \bibfield  {author} {\bibinfo {author} {\bibfnamefont {M.}~\bibnamefont
  {Baiesi}}, \bibinfo {author} {\bibfnamefont {C.}~\bibnamefont {Maes}},\ and\
  \bibinfo {author} {\bibfnamefont {B.}~\bibnamefont {Wynants}},\ }\href
  {https://doi.org/10.1103/PhysRevLett.103.010602} {\bibfield  {journal}
  {\bibinfo  {journal} {Physical Review Letters}\ }\textbf {\bibinfo {volume}
  {103}},\ \bibinfo {pages} {010602} (\bibinfo {year} {2009})}\BibitemShut
  {NoStop}%
\bibitem [{\citenamefont {Martin}\ \emph {et~al.}(2001)\citenamefont {Martin},
  \citenamefont {Hudspeth},\ and\ \citenamefont
  {Jülicher}}]{martin_comparison_2001}%
  \BibitemOpen
  \bibfield  {author} {\bibinfo {author} {\bibfnamefont {P.}~\bibnamefont
  {Martin}}, \bibinfo {author} {\bibfnamefont {A.~J.}\ \bibnamefont
  {Hudspeth}},\ and\ \bibinfo {author} {\bibfnamefont {F.}~\bibnamefont
  {Jülicher}},\ }\href {https://doi.org/10.1073/pnas.251530598} {\bibfield
  {journal} {\bibinfo  {journal} {Proceedings of the National Academy of
  Sciences}\ }\textbf {\bibinfo {volume} {98}},\ \bibinfo {pages}
  {14380–14385} (\bibinfo {year} {2001})}\BibitemShut {NoStop}%
\bibitem [{\citenamefont {Harada}\ and\ \citenamefont
  {Sasa}(2005)}]{harada_equality_2005}%
  \BibitemOpen
  \bibfield  {author} {\bibinfo {author} {\bibfnamefont {T.}~\bibnamefont
  {Harada}}\ and\ \bibinfo {author} {\bibfnamefont {S.-i.}\ \bibnamefont
  {Sasa}},\ }\href {https://doi.org/10.1103/PhysRevLett.95.130602} {\bibfield
  {journal} {\bibinfo  {journal} {Physical Review Letters}\ }\textbf {\bibinfo
  {volume} {95}},\ \bibinfo {pages} {130602} (\bibinfo {year}
  {2005})}\BibitemShut {NoStop}%
\bibitem [{\citenamefont {Toyabe}\ \emph {et~al.}(2010)\citenamefont {Toyabe},
  \citenamefont {Okamoto}, \citenamefont {Watanabe-Nakayama}, \citenamefont
  {Taketani}, \citenamefont {Kudo},\ and\ \citenamefont
  {Muneyuki}}]{toyabe_nonequilibrium_2010}%
  \BibitemOpen
  \bibfield  {author} {\bibinfo {author} {\bibfnamefont {S.}~\bibnamefont
  {Toyabe}}, \bibinfo {author} {\bibfnamefont {T.}~\bibnamefont {Okamoto}},
  \bibinfo {author} {\bibfnamefont {T.}~\bibnamefont {Watanabe-Nakayama}},
  \bibinfo {author} {\bibfnamefont {H.}~\bibnamefont {Taketani}}, \bibinfo
  {author} {\bibfnamefont {S.}~\bibnamefont {Kudo}},\ and\ \bibinfo {author}
  {\bibfnamefont {E.}~\bibnamefont {Muneyuki}},\ }\href
  {https://doi.org/10.1103/PhysRevLett.104.198103} {\bibfield  {journal}
  {\bibinfo  {journal} {Physical Review Letters}\ }\textbf {\bibinfo {volume}
  {104}},\ \bibinfo {pages} {198103} (\bibinfo {year} {2010})}\BibitemShut
  {NoStop}%
\bibitem [{\citenamefont {Krüger}\ and\ \citenamefont
  {Fuchs}(2009)}]{kruger_fluctuation_2009}%
  \BibitemOpen
  \bibfield  {author} {\bibinfo {author} {\bibfnamefont {M.}~\bibnamefont
  {Krüger}}\ and\ \bibinfo {author} {\bibfnamefont {M.}~\bibnamefont
  {Fuchs}},\ }\href {https://doi.org/10.1103/PhysRevLett.102.135701} {\bibfield
   {journal} {\bibinfo  {journal} {Physical Review Letters}\ }\textbf {\bibinfo
  {volume} {102}},\ \bibinfo {pages} {135701} (\bibinfo {year}
  {2009})}\BibitemShut {NoStop}%
\bibitem [{\citenamefont {Ahmed}\ \emph {et~al.}(2018)\citenamefont {Ahmed},
  \citenamefont {Fodor}, \citenamefont {Almonacid}, \citenamefont {Bussonnier},
  \citenamefont {Verlhac}, \citenamefont {Gov}, \citenamefont {Visco},
  \citenamefont {Wijland},\ and\ \citenamefont {Betz}}]{ahmed_active_2018}%
  \BibitemOpen
  \bibfield  {author} {\bibinfo {author} {\bibfnamefont {W.~W.}\ \bibnamefont
  {Ahmed}}, \bibinfo {author} {\bibfnamefont {{\'{E}}.}~\bibnamefont {Fodor}},
  \bibinfo {author} {\bibfnamefont {M.}~\bibnamefont {Almonacid}}, \bibinfo
  {author} {\bibfnamefont {M.}~\bibnamefont {Bussonnier}}, \bibinfo {author}
  {\bibfnamefont {M.-H.}\ \bibnamefont {Verlhac}}, \bibinfo {author}
  {\bibfnamefont {N.}~\bibnamefont {Gov}}, \bibinfo {author} {\bibfnamefont
  {P.}~\bibnamefont {Visco}}, \bibinfo {author} {\bibfnamefont {F.~v.}\
  \bibnamefont {Wijland}},\ and\ \bibinfo {author} {\bibfnamefont
  {T.}~\bibnamefont {Betz}},\ }\href
  {https://doi.org/10.1016/j.bpj.2018.02.009} {\bibfield  {journal} {\bibinfo
  {journal} {Biophysical Journal}\ }\textbf {\bibinfo {volume} {114}},\
  \bibinfo {pages} {1667–1679} (\bibinfo {year} {2018})}\BibitemShut
  {NoStop}%
\bibitem [{\citenamefont {Muenker}\ \emph {et~al.}(2022)\citenamefont
  {Muenker}, \citenamefont {Knotz}, \citenamefont {Krüger},\ and\
  \citenamefont {Betz}}]{muenker_onsager_2022}%
  \BibitemOpen
  \bibfield  {author} {\bibinfo {author} {\bibfnamefont {T.~M.}\ \bibnamefont
  {Muenker}}, \bibinfo {author} {\bibfnamefont {G.}~\bibnamefont {Knotz}},
  \bibinfo {author} {\bibfnamefont {M.}~\bibnamefont {Krüger}},\ and\ \bibinfo
  {author} {\bibfnamefont {T.}~\bibnamefont {Betz}},\ }\bibfield  {journal}
  {\bibinfo  {journal} {bioRxiv}\ }\href
  {https://doi.org/10.1101/2022.05.15.491928} {10.1101/2022.05.15.491928}
  (\bibinfo {year} {2022})\BibitemShut {NoStop}%
\bibitem [{\citenamefont {Jarzynski}(1997)}]{jarzynski_nonequilibrium_1997}%
  \BibitemOpen
  \bibfield  {author} {\bibinfo {author} {\bibfnamefont {C.}~\bibnamefont
  {Jarzynski}},\ }\href {https://doi.org/10.1103/PhysRevLett.78.2690}
  {\bibfield  {journal} {\bibinfo  {journal} {Physical Review Letters}\
  }\textbf {\bibinfo {volume} {78}},\ \bibinfo {pages} {2690–2693} (\bibinfo
  {year} {1997})}\BibitemShut {NoStop}%
\bibitem [{\citenamefont {Crooks}(1999)}]{Crooks99}%
  \BibitemOpen
  \bibfield  {author} {\bibinfo {author} {\bibfnamefont {G.~E.}\ \bibnamefont
  {Crooks}},\ }\href {https://doi.org/10.1103/PhysRevE.60.2721} {\bibfield
  {journal} {\bibinfo  {journal} {Phys. Rev. E}\ }\textbf {\bibinfo {volume}
  {60}},\ \bibinfo {pages} {2721} (\bibinfo {year} {1999})}\BibitemShut
  {NoStop}%
\bibitem [{\citenamefont {Speck}\ and\ \citenamefont
  {Seifert}(2007)}]{Speck_2007}%
  \BibitemOpen
  \bibfield  {author} {\bibinfo {author} {\bibfnamefont {T.}~\bibnamefont
  {Speck}}\ and\ \bibinfo {author} {\bibfnamefont {U.}~\bibnamefont
  {Seifert}},\ }\href {https://doi.org/10.1088/1742-5468/2007/09/L09002}
  {\bibfield  {journal} {\bibinfo  {journal} {Journal of Statistical Mechanics:
  Theory and Experiment}\ }\textbf {\bibinfo {volume} {2007}},\ \bibinfo
  {pages} {L09002} (\bibinfo {year} {2007})}\BibitemShut {NoStop}%
\bibitem [{\citenamefont {Seifert}(2012)}]{Seifert2012}%
  \BibitemOpen
  \bibfield  {author} {\bibinfo {author} {\bibfnamefont {U.}~\bibnamefont
  {Seifert}},\ }\href {https://doi.org/10.1088/0034-4885/75/12/126001}
  {\bibfield  {journal} {\bibinfo  {journal} {Reports on Progress in Physics}\
  }\textbf {\bibinfo {volume} {75}},\ \bibinfo {pages} {126001} (\bibinfo
  {year} {2012})}\BibitemShut {NoStop}%
\bibitem [{\citenamefont {Barato}\ and\ \citenamefont
  {Seifert}(2015)}]{barato_thermodynamic_2015}%
  \BibitemOpen
  \bibfield  {author} {\bibinfo {author} {\bibfnamefont {A.~C.}\ \bibnamefont
  {Barato}}\ and\ \bibinfo {author} {\bibfnamefont {U.}~\bibnamefont
  {Seifert}},\ }\href {https://doi.org/10.1103/PhysRevLett.114.158101}
  {\bibfield  {journal} {\bibinfo  {journal} {Physical Review Letters}\
  }\textbf {\bibinfo {volume} {114}},\ \bibinfo {pages} {158101} (\bibinfo
  {year} {2015})}\BibitemShut {NoStop}%
\bibitem [{\citenamefont {Horowitz}\ and\ \citenamefont
  {Gingrich}(2020)}]{Horowitz2020}%
  \BibitemOpen
  \bibfield  {author} {\bibinfo {author} {\bibfnamefont {J.~M.}\ \bibnamefont
  {Horowitz}}\ and\ \bibinfo {author} {\bibfnamefont {T.~R.}\ \bibnamefont
  {Gingrich}},\ }\href {https://doi.org/10.1038/s41567-019-0702-6} {\bibfield
  {journal} {\bibinfo  {journal} {Nature Physics}\ }\textbf {\bibinfo {volume}
  {16}},\ \bibinfo {pages} {15–20} (\bibinfo {year} {2020})}\BibitemShut
  {NoStop}%
\bibitem [{\citenamefont {Li}\ \emph {et~al.}(2019)\citenamefont {Li},
  \citenamefont {Horowitz}, \citenamefont {Gingrich},\ and\ \citenamefont
  {Fakhri}}]{li_quantifying_2019}%
  \BibitemOpen
  \bibfield  {author} {\bibinfo {author} {\bibfnamefont {J.}~\bibnamefont
  {Li}}, \bibinfo {author} {\bibfnamefont {J.~M.}\ \bibnamefont {Horowitz}},
  \bibinfo {author} {\bibfnamefont {T.~R.}\ \bibnamefont {Gingrich}},\ and\
  \bibinfo {author} {\bibfnamefont {N.}~\bibnamefont {Fakhri}},\ }\href
  {https://doi.org/10.1038/s41467-019-09631-x} {\bibfield  {journal} {\bibinfo
  {journal} {Nature Communications}\ }\textbf {\bibinfo {volume} {10}},\
  \bibinfo {pages} {1666} (\bibinfo {year} {2019})}\BibitemShut {NoStop}%
\bibitem [{\citenamefont {{Van Vu}}\ \emph {et~al.}(2020)\citenamefont {{Van
  Vu}}, \citenamefont {Vo},\ and\ \citenamefont
  {Hasegawa}}]{van_vu_entropy_2020}%
  \BibitemOpen
  \bibfield  {author} {\bibinfo {author} {\bibfnamefont {T.}~\bibnamefont {{Van
  Vu}}}, \bibinfo {author} {\bibfnamefont {V.~T.}\ \bibnamefont {Vo}},\ and\
  \bibinfo {author} {\bibfnamefont {Y.}~\bibnamefont {Hasegawa}},\ }\href
  {https://doi.org/10.1103/PhysRevE.101.042138} {\bibfield  {journal} {\bibinfo
   {journal} {Physical Review E}\ }\textbf {\bibinfo {volume} {101}},\ \bibinfo
  {pages} {042138} (\bibinfo {year} {2020})}\BibitemShut {NoStop}%
\bibitem [{\citenamefont {Dechant}\ and\ \citenamefont
  {Sasa}(2021)}]{dechant_improving_2021}%
  \BibitemOpen
  \bibfield  {author} {\bibinfo {author} {\bibfnamefont {A.}~\bibnamefont
  {Dechant}}\ and\ \bibinfo {author} {\bibfnamefont {S.-i.}\ \bibnamefont
  {Sasa}},\ }\href {https://doi.org/10.1103/PhysRevX.11.041061} {\bibfield
  {journal} {\bibinfo  {journal} {Physical Review X}\ }\textbf {\bibinfo
  {volume} {11}},\ \bibinfo {pages} {041061} (\bibinfo {year}
  {2021})}\BibitemShut {NoStop}%
\bibitem [{\citenamefont {Pietzonka}(2022)}]{pietzonka_classical_2022}%
  \BibitemOpen
  \bibfield  {author} {\bibinfo {author} {\bibfnamefont {P.}~\bibnamefont
  {Pietzonka}},\ }\href {https://doi.org/10.1103/PhysRevLett.128.130606}
  {\bibfield  {journal} {\bibinfo  {journal} {Physical Review Letters}\
  }\textbf {\bibinfo {volume} {128}},\ \bibinfo {pages} {130606} (\bibinfo
  {year} {2022})}\BibitemShut {NoStop}%
\bibitem [{\citenamefont {Cao}\ \emph {et~al.}(2022)\citenamefont {Cao},
  \citenamefont {Su}, \citenamefont {Jiang},\ and\ \citenamefont
  {Hou}}]{cao_effective_2022}%
  \BibitemOpen
  \bibfield  {author} {\bibinfo {author} {\bibfnamefont {Z.}~\bibnamefont
  {Cao}}, \bibinfo {author} {\bibfnamefont {J.}~\bibnamefont {Su}}, \bibinfo
  {author} {\bibfnamefont {H.}~\bibnamefont {Jiang}},\ and\ \bibinfo {author}
  {\bibfnamefont {Z.}~\bibnamefont {Hou}},\ }\href
  {https://doi.org/10.1063/5.0094211} {\bibfield  {journal} {\bibinfo
  {journal} {Physics of Fluids}\ }\textbf {\bibinfo {volume} {34}},\ \bibinfo
  {pages} {053310} (\bibinfo {year} {2022})}\BibitemShut {NoStop}%
\bibitem [{\citenamefont {Koyuk}\ and\ \citenamefont
  {Seifert}(2022)}]{koyuk_thermodynamic_2022}%
  \BibitemOpen
  \bibfield  {author} {\bibinfo {author} {\bibfnamefont {T.}~\bibnamefont
  {Koyuk}}\ and\ \bibinfo {author} {\bibfnamefont {U.}~\bibnamefont
  {Seifert}},\ }\href {https://doi.org/10.1103/PhysRevLett.129.210603}
  {\bibfield  {journal} {\bibinfo  {journal} {Physical Review Letters}\
  }\textbf {\bibinfo {volume} {129}},\ \bibinfo {pages} {210603} (\bibinfo
  {year} {2022})}\BibitemShut {NoStop}%
\bibitem [{\citenamefont {Dechant}\ and\ \citenamefont
  {Sasa}(2018)}]{dechant_current_2018}%
  \BibitemOpen
  \bibfield  {author} {\bibinfo {author} {\bibfnamefont {A.}~\bibnamefont
  {Dechant}}\ and\ \bibinfo {author} {\bibfnamefont {S.-i.}\ \bibnamefont
  {Sasa}},\ }\href {https://doi.org/10.1088/1742-5468/aac91a} {\bibfield
  {journal} {\bibinfo  {journal} {Journal of Statistical Mechanics: Theory and
  Experiment}\ }\textbf {\bibinfo {volume} {2018}},\ \bibinfo {pages} {063209}
  (\bibinfo {year} {2018})}\BibitemShut {NoStop}%
\bibitem [{\citenamefont {Hasegawa}\ and\ \citenamefont {{Van
  Vu}}(2019)}]{hasegawa_fluctuation_2019}%
  \BibitemOpen
  \bibfield  {author} {\bibinfo {author} {\bibfnamefont {Y.}~\bibnamefont
  {Hasegawa}}\ and\ \bibinfo {author} {\bibfnamefont {T.}~\bibnamefont {{Van
  Vu}}},\ }\href {https://doi.org/10.1103/PhysRevLett.123.110602} {\bibfield
  {journal} {\bibinfo  {journal} {Physical Review Letters}\ }\textbf {\bibinfo
  {volume} {123}},\ \bibinfo {pages} {110602} (\bibinfo {year}
  {2019})}\BibitemShut {NoStop}%
\bibitem [{\citenamefont {Dechant}\ and\ \citenamefont
  {Sasa}(2020)}]{dechant_fluctuationresponse_2020}%
  \BibitemOpen
  \bibfield  {author} {\bibinfo {author} {\bibfnamefont {A.}~\bibnamefont
  {Dechant}}\ and\ \bibinfo {author} {\bibfnamefont {S.-i.}\ \bibnamefont
  {Sasa}},\ }\href {https://doi.org/10.1073/pnas.1918386117} {\bibfield
  {journal} {\bibinfo  {journal} {Proceedings of the National Academy of
  Sciences}\ }\textbf {\bibinfo {volume} {117}},\ \bibinfo {pages}
  {6430–6436} (\bibinfo {year} {2020})}\BibitemShut {NoStop}%
\bibitem [{\citenamefont {Ziyin}\ and\ \citenamefont
  {Ueda}(2023)}]{ziyin_universal_2023}%
  \BibitemOpen
  \bibfield  {author} {\bibinfo {author} {\bibfnamefont {L.}~\bibnamefont
  {Ziyin}}\ and\ \bibinfo {author} {\bibfnamefont {M.}~\bibnamefont {Ueda}},\
  }\href {https://doi.org/10.1103/PhysRevResearch.5.013039} {\bibfield
  {journal} {\bibinfo  {journal} {Physical Review Research}\ }\textbf {\bibinfo
  {volume} {5}},\ \bibinfo {pages} {013039} (\bibinfo {year}
  {2023})}\BibitemShut {NoStop}%
\bibitem [{\citenamefont {Altland}\ and\ \citenamefont
  {Simons}(2010)}]{altland_condensed_2010}%
  \BibitemOpen
  \bibfield  {author} {\bibinfo {author} {\bibfnamefont {A.}~\bibnamefont
  {Altland}}\ and\ \bibinfo {author} {\bibfnamefont {B.~D.}\ \bibnamefont
  {Simons}},\ }\href {https://doi.org/10.1017/CBO9780511789984} {\emph
  {\bibinfo {title} {Condensed {Matter} {Field} {Theory}}}},\ \bibinfo
  {edition} {2nd}\ ed.\ (\bibinfo  {publisher} {Cambridge University Press},\
  \bibinfo {year} {2010})\BibitemShut {NoStop}%
\bibitem [{\citenamefont {García-García}\ \emph {et~al.}(2010)\citenamefont
  {García-García}, \citenamefont {Domínguez}, \citenamefont {Lecomte},\ and\
  \citenamefont {Kolton}}]{garcia-garcia_unifying_2010}%
  \BibitemOpen
  \bibfield  {author} {\bibinfo {author} {\bibfnamefont {R.}~\bibnamefont
  {García-García}}, \bibinfo {author} {\bibfnamefont {D.}~\bibnamefont
  {Domínguez}}, \bibinfo {author} {\bibfnamefont {V.}~\bibnamefont
  {Lecomte}},\ and\ \bibinfo {author} {\bibfnamefont {A.~B.}\ \bibnamefont
  {Kolton}},\ }\href {https://doi.org/10.1103/PhysRevE.82.030104} {\bibfield
  {journal} {\bibinfo  {journal} {Physical Review E}\ }\textbf {\bibinfo
  {volume} {82}},\ \bibinfo {pages} {030104} (\bibinfo {year}
  {2010})}\BibitemShut {NoStop}%
\bibitem [{\citenamefont {Spinney}\ and\ \citenamefont
  {Ford}(2012)}]{spinney_entropy_2012}%
  \BibitemOpen
  \bibfield  {author} {\bibinfo {author} {\bibfnamefont {R.~E.}\ \bibnamefont
  {Spinney}}\ and\ \bibinfo {author} {\bibfnamefont {I.~J.}\ \bibnamefont
  {Ford}},\ }\href {https://doi.org/10.1103/PhysRevE.85.051113} {\bibfield
  {journal} {\bibinfo  {journal} {Physical Review E}\ }\textbf {\bibinfo
  {volume} {85}},\ \bibinfo {pages} {051113} (\bibinfo {year}
  {2012})}\BibitemShut {NoStop}%
\bibitem [{\citenamefont {Harris}\ and\ \citenamefont
  {Schütz}(2007)}]{harris_fluctuation_2007}%
  \BibitemOpen
  \bibfield  {author} {\bibinfo {author} {\bibfnamefont {R.~J.}\ \bibnamefont
  {Harris}}\ and\ \bibinfo {author} {\bibfnamefont {G.~M.}\ \bibnamefont
  {Schütz}},\ }\href {https://doi.org/10.1088/1742-5468/2007/07/P07020}
  {\bibfield  {journal} {\bibinfo  {journal} {Journal of Statistical Mechanics:
  Theory and Experiment}\ }\textbf {\bibinfo {volume} {2007}},\ \bibinfo
  {pages} {P07020–P07020} (\bibinfo {year} {2007})}\BibitemShut {NoStop}%
\bibitem [{\citenamefont {Zia}\ and\ \citenamefont
  {Schmittmann}(2007)}]{zia_probability_2007}%
  \BibitemOpen
  \bibfield  {author} {\bibinfo {author} {\bibfnamefont {R.~K.~P.}\
  \bibnamefont {Zia}}\ and\ \bibinfo {author} {\bibfnamefont {B.}~\bibnamefont
  {Schmittmann}},\ }\href {https://doi.org/10.1088/1742-5468/2007/07/P07012}
  {\bibfield  {journal} {\bibinfo  {journal} {Journal of Statistical Mechanics:
  Theory and Experiment}\ }\textbf {\bibinfo {volume} {2007}},\ \bibinfo
  {pages} {P07012–P07012} (\bibinfo {year} {2007})}\BibitemShut {NoStop}%
\bibitem [{\citenamefont {Maes}\ and\ \citenamefont
  {Netočný}(2003)}]{Maes2003}%
  \BibitemOpen
  \bibfield  {author} {\bibinfo {author} {\bibfnamefont {C.}~\bibnamefont
  {Maes}}\ and\ \bibinfo {author} {\bibfnamefont {K.}~\bibnamefont
  {Netočný}},\ }\href {https://doi.org/10.1023/A:1021026930129} {\bibfield
  {journal} {\bibinfo  {journal} {Journal of Statistical Physics}\ }\textbf
  {\bibinfo {volume} {110}},\ \bibinfo {pages} {269–310} (\bibinfo {year}
  {2003})}\BibitemShut {NoStop}%
\bibitem [{\citenamefont {Seifert}(2005)}]{seifert_entropy_2005}%
  \BibitemOpen
  \bibfield  {author} {\bibinfo {author} {\bibfnamefont {U.}~\bibnamefont
  {Seifert}},\ }\href {https://doi.org/10.1103/PhysRevLett.95.040602}
  {\bibfield  {journal} {\bibinfo  {journal} {Physical Review Letters}\
  }\textbf {\bibinfo {volume} {95}},\ \bibinfo {pages} {040602} (\bibinfo
  {year} {2005})}\BibitemShut {NoStop}%
\bibitem [{\citenamefont {Fischer}\ \emph {et~al.}(2020)\citenamefont
  {Fischer}, \citenamefont {Chun},\ and\ \citenamefont
  {Seifert}}]{fischer_free_2020}%
  \BibitemOpen
  \bibfield  {author} {\bibinfo {author} {\bibfnamefont {L.~P.}\ \bibnamefont
  {Fischer}}, \bibinfo {author} {\bibfnamefont {H.-M.}\ \bibnamefont {Chun}},\
  and\ \bibinfo {author} {\bibfnamefont {U.}~\bibnamefont {Seifert}},\ }\href
  {https://doi.org/10.1103/PhysRevE.102.012120} {\bibfield  {journal} {\bibinfo
   {journal} {Physical Review E}\ }\textbf {\bibinfo {volume} {102}},\ \bibinfo
  {pages} {012120} (\bibinfo {year} {2020})}\BibitemShut {NoStop}%
\bibitem [{Note1()}]{Note1}%
  \BibitemOpen
  \bibinfo {note} {The entropy defined in Eq.~\protect \eqref {eq:def-ent}
  corresponds to the total change in entropy in overdamped stationary systems.
  In underdamped or non-stationary systems the boundary terms differ \cite
  {fischer_free_2020}.}\BibitemShut {Stop}%
\bibitem [{\citenamefont {Jensen}(1906)}]{jensen_sur_1906}%
  \BibitemOpen
  \bibfield  {author} {\bibinfo {author} {\bibfnamefont {J.~L. W.~V.}\
  \bibnamefont {Jensen}},\ }\href {https://doi.org/10.1007/BF02418571}
  {\bibfield  {journal} {\bibinfo  {journal} {Acta Mathematica}\ }\textbf
  {\bibinfo {volume} {30}},\ \bibinfo {pages} {175–193} (\bibinfo {year}
  {1906})}\BibitemShut {NoStop}%
\bibitem [{\citenamefont {Durrett}(2019)}]{durrett_probabilitytheory_2019}%
  \BibitemOpen
  \bibfield  {author} {\bibinfo {author} {\bibfnamefont {R.}~\bibnamefont
  {Durrett}},\ }\href@noop {} {\emph {\bibinfo {title} {Probability:{Theory}
  and {Examples}}}},\ \bibinfo {edition} {5th}\ ed.,\ Cambridge {Series} in
  {Statistical} and {Probabilistic} {Mathematics}\ (\bibinfo  {publisher}
  {Cambridge University Press},\ \bibinfo {address} {Cambridge, England},\
  \bibinfo {year} {2019})\BibitemShut {NoStop}%
\bibitem [{Note2()}]{Note2}%
  \BibitemOpen
  \bibinfo {note} {A similar relation was derived in Ref.~\cite
  {merhav_statistical_2010}, however, with the left hand side always
  negative.}\BibitemShut {Stop}%
\bibitem [{\citenamefont {Shiraishi}(2021)}]{shiraishi_optimal_2021}%
  \BibitemOpen
  \bibfield  {author} {\bibinfo {author} {\bibfnamefont {N.}~\bibnamefont
  {Shiraishi}},\ }\href {https://doi.org/10.1007/s10955-021-02829-8} {\bibfield
   {journal} {\bibinfo  {journal} {Journal of Statistical Physics}\ }\textbf
  {\bibinfo {volume} {185}},\ \bibinfo {pages} {19} (\bibinfo {year}
  {2021})}\BibitemShut {NoStop}%
\bibitem [{\citenamefont {Dieball}\ and\ \citenamefont
  {Godec}(2023)}]{dieball_direct_2023}%
  \BibitemOpen
  \bibfield  {author} {\bibinfo {author} {\bibfnamefont {C.}~\bibnamefont
  {Dieball}}\ and\ \bibinfo {author} {\bibfnamefont {A.}~\bibnamefont
  {Godec}},\ }\href {https://doi.org/10.1103/PhysRevLett.130.087101} {\bibfield
   {journal} {\bibinfo  {journal} {Physical Review Letters}\ }\textbf {\bibinfo
  {volume} {130}},\ \bibinfo {pages} {087101} (\bibinfo {year}
  {2023})}\BibitemShut {NoStop}%
\bibitem [{Note3()}]{Note3}%
  \BibitemOpen
  \bibinfo {note} {Eq.~\protect \eqref {eq:sigs} holds also for $-O_a$ and thus
  for $|\left \langle O_a \right \rangle |$.}\BibitemShut {Stop}%
\bibitem [{Note4()}]{Note4}%
  \BibitemOpen
  \bibinfo {note} {The terms with $p[\omega ] = p[\theta \omega ]$ cancel in
  the sum in Eq.~\protect \eqref {eq:sigs} due to $O_a[\omega ]=-O_a[\theta
  \omega ]$, and $O_a[\omega ]$ can be chosen arbitrarily in these
  cases.}\BibitemShut {Stop}%
\bibitem [{\citenamefont {Seifert}(2019)}]{seifert_stochastic_2019}%
  \BibitemOpen
  \bibfield  {author} {\bibinfo {author} {\bibfnamefont {U.}~\bibnamefont
  {Seifert}},\ }\href
  {https://doi.org/10.1146/annurev-conmatphys-031218-013554} {\bibfield
  {journal} {\bibinfo  {journal} {Annual Review of Condensed Matter Physics}\
  }\textbf {\bibinfo {volume} {10}},\ \bibinfo {pages} {171–192} (\bibinfo
  {year} {2019})}\BibitemShut {NoStop}%
\bibitem [{\citenamefont {Arratia}\ and\ \citenamefont
  {Gordon}(1989)}]{arratia_tutorial_1989}%
  \BibitemOpen
  \bibfield  {author} {\bibinfo {author} {\bibfnamefont {R.}~\bibnamefont
  {Arratia}}\ and\ \bibinfo {author} {\bibfnamefont {L.}~\bibnamefont
  {Gordon}},\ }\href {https://doi.org/10.1007/BF02458840} {\bibfield  {journal}
  {\bibinfo  {journal} {Bulletin of Mathematical Biology}\ }\textbf {\bibinfo
  {volume} {51}},\ \bibinfo {pages} {125–131} (\bibinfo {year}
  {1989})}\BibitemShut {NoStop}%
\bibitem [{\citenamefont {Esposito}\ \emph {et~al.}(2009)\citenamefont
  {Esposito}, \citenamefont {Harbola},\ and\ \citenamefont
  {Mukamel}}]{esposito_nonequilibrium_2009}%
  \BibitemOpen
  \bibfield  {author} {\bibinfo {author} {\bibfnamefont {M.}~\bibnamefont
  {Esposito}}, \bibinfo {author} {\bibfnamefont {U.}~\bibnamefont {Harbola}},\
  and\ \bibinfo {author} {\bibfnamefont {S.}~\bibnamefont {Mukamel}},\ }\href
  {https://doi.org/10.1103/RevModPhys.81.1665} {\bibfield  {journal} {\bibinfo
  {journal} {Reviews of Modern Physics}\ }\textbf {\bibinfo {volume} {81}},\
  \bibinfo {pages} {1665–1702} (\bibinfo {year} {2009})}\BibitemShut
  {NoStop}%
\bibitem [{\citenamefont {Landi}\ and\ \citenamefont
  {Paternostro}(2021)}]{landi_irreversible_2021}%
  \BibitemOpen
  \bibfield  {author} {\bibinfo {author} {\bibfnamefont {G.~T.}\ \bibnamefont
  {Landi}}\ and\ \bibinfo {author} {\bibfnamefont {M.}~\bibnamefont
  {Paternostro}},\ }\href {https://doi.org/10.1103/RevModPhys.93.035008}
  {\bibfield  {journal} {\bibinfo  {journal} {Reviews of Modern Physics}\
  }\textbf {\bibinfo {volume} {93}},\ \bibinfo {pages} {035008} (\bibinfo
  {year} {2021})}\BibitemShut {NoStop}%
\bibitem [{\citenamefont {Aron}\ \emph {et~al.}(2018)\citenamefont {Aron},
  \citenamefont {Biroli},\ and\ \citenamefont {Cugliandolo}}]{aron_non_2018}%
  \BibitemOpen
  \bibfield  {author} {\bibinfo {author} {\bibfnamefont {C.}~\bibnamefont
  {Aron}}, \bibinfo {author} {\bibfnamefont {G.}~\bibnamefont {Biroli}},\ and\
  \bibinfo {author} {\bibfnamefont {L.}~\bibnamefont {Cugliandolo}},\ }\href
  {https://doi.org/10.21468/SciPostPhys.4.1.008} {\bibfield  {journal}
  {\bibinfo  {journal} {SciPost Physics}\ }\textbf {\bibinfo {volume} {4}},\
  \bibinfo {pages} {008} (\bibinfo {year} {2018})}\BibitemShut {NoStop}%
\bibitem [{\citenamefont {Merhav}\ and\ \citenamefont
  {Kafri}(2010)}]{merhav_statistical_2010}%
  \BibitemOpen
  \bibfield  {author} {\bibinfo {author} {\bibfnamefont {N.}~\bibnamefont
  {Merhav}}\ and\ \bibinfo {author} {\bibfnamefont {Y.}~\bibnamefont {Kafri}},\
  }\href {https://doi.org/10.1088/1742-5468/2010/12/P12022} {\bibfield
  {journal} {\bibinfo  {journal} {Journal of Statistical Mechanics: Theory and
  Experiment}\ }\textbf {\bibinfo {volume} {2010}},\ \bibinfo {pages} {P12022}
  (\bibinfo {year} {2010})}\BibitemShut {NoStop}%
\end{thebibliography}%
\bibliographystyle{apsrev4-2}

\end{document}